# A Comparative Study of AI-based Intrusion Detection Techniques in Critical Infrastructures

SAFA OTOUM, College of Technological Innovation, Zayed University, UAE
BURAK KANTARCI AND HUSSEIN MOUFTAH, University of Ottawa

Volunteer computing uses Internet-connected devices (laptops, PCs, smart devices,etc.), in which their owners volunteer them as storage and computing power resources, has become an essential mechanism for resource management in numerous applications. The growth of the volume and variety of data traffic in the Internet leads to concerns on the robustness of cyberphysical systems especially for critical infrastructures. Therefore, the implementation of an efficient Intrusion Detection System for gathering such sensory data has gained vital importance. In this paper, we present a comparative study of Artificial Intelligence (AI)-driven intrusion detection systems for wirelessly connected sensors that track crucial applications. Specifically, we present an in-depth analysis of the use of machine learning, deep learning and reinforcement learning solutions to recognize intrusive behavior in the collected traffic. We evaluate the proposed mechanisms by using KDD'99 as real attack data-set in our simulations. Results present the performance metrics for three different IDSs namely the Adaptively Supervised and Clustered Hybrid IDS (ASCH-IDS), Restricted Boltzmann Machine-based Clustered IDS (RBC-IDS) and Q-learning based IDS (QL-IDS) to detect malicious behaviors. We also present the performance of different reinforcement learning techniques such as State-Action-Reward-State-Action Learning (SARSA) and the Temporal Difference learning (TD). Through simulations, we show that QL-IDS performs with 100% detection rate while SARSA-IDS and TD-IDS perform at the order of 99.5%.

## 1 INTRODUCTION

Smart cities exploit the flexibility, self-deployment and low-cost benefits of wireless sensors to monitor their critical infrastructures [1]. Monitoring critical infrastructures involves various types of sensors under different circumstances [2][3]. Efficiency and reliability of critical infrastructures such as medical infrastructure can be met by using reliable and secure data aggregation and transmission [4] [5]. Sensors and the communication lines between them are vulnerable to numerous intruders that may interrupt and manipulate with the communication and transmitted data [6][7]. IDS is an essential component for network security to detect different types of intrusions. At the very high

Authors' addresses: Safa Otoum, Safa.Otoum@zu.ac.ae, College of Technological Innovation, Zayed University, UAE, Abu Dhabi, UAE; Burak Kantarci and Hussein Mouftah, Burak.Kantarci;Mouftah@uottawa.ca, University of Ottawa, 800 King Edward Ave, Ottawa, ON, K1N6N5.

level, the validity of any IDS is determined by its ability of raising an alarm with the detection of any malicious node that exhibits intrusive behavior [8]. IDSs are categorized in two sets: Anomaly-based IDS and Signature-based IDS. The former aims to detect abnormal traffic patterns that deviate from the normal behavior. To this end, it creates profiles of the features to seize the needed patterns. The variance between the extracted patterns and noticed activities may lead to an alarm. Thus, ideally Anomaly-based IDSs are able to detect unknown attacks. However, It is worth noting that most IDS systems still suffer from False Positive (FP) decisions, i.e., a non-malicious activity's being marked as an intrusive event [6]. Apart from the anomaly-based IDS, signature IDSs uses the rule complement mechanism to detect malicious behavior by comparing system activities with specific rules. An intrusive event is detected when the noticed activities match a known malicious pattern. The risk of solely adopting a signature-based IDS is that its possible low precision in the likelihood of an unknown attack [6][8]. In our previous work, we proposed several artificial intelligence (AI)-based IDS methods for WSNs monitoring critical infrastructures [8][7]. All of these approaches have the following aim in common: signature detection and anomaly detection which can be combined in an IDS by following an adaptive supervision procedure. In this paper, we provide a comparative study of intrusion detection in critical infrastructures which compares the usage of machine learning (random forest and E-DBSCAN), deep learning (restricted Boltzmann machine) and reinforcement learning approaches (Q-learning). We also develop our work presented in [6][8][7] by introducing Q-learning, SARSA-learning and TD-learning based reinforcement procedures to our previously proposed IDS. The main contributions of this work are mentioned as follows.

(1) This work directly target the volunteer computing by applying the previously proposed IDSs to the distributed data and evaluating the IDS performance.
(2) It presents an intrusion detection mechanism, in which reinforcement learning using Q-learning, SARSA learning and TD learning methods are introduced to estimate the IDS performance.
(3) For the first time, we present a comprehensive and comparative study of IDS evaluation using machine, deep and reinforcement learning methods.

## 2 MOTIVATIONS

WSNs are major technology in critical infrastructures, especially in monitoring and control processes, due to their self-deployment, flexibility and low-cost benefits. Monitoring critical infrastructures involve numerous sensors under various phenomena. In most of critical infrastructures, WSNs are deployed in an open nature environment, thus, nodes could be vulnerable to different attacks which can manipulate the aggregated sensory data. The communication lines are also vulnerable to numerous intruders that may interrupt and manipulate with the communication and transmitted data. WSNs need to provide an acceptable level of security to achieve attacks free environments, reliable and secure data aggregation and transmission [4]. Intrusion Detection System (IDS) is an essential component for network security especially for the sensor networks to detect different types of intrusions. At the very high level, the validity of any IDS is determined by its ability to raise an alarm with the detection of any malicious node that exhibits intrusive behavior [8].

The rest of the paper is organized as follows. Section 3 layouts this study's related works. Section 4 introduces the IDS topology settings. Section 5 describes the methods used to test the previously proposed IDS. Experimental and results analysis are presented in section 6. Finally, we present the conclusion in Section 7.

## 3 RELATED WORKS

Integration of machine learning with IDS have been proposed by many researchers. Some of them tackled the unknown attacks [9] while others tackled the known ones [10]. Lately, deep learning based techniques had been used in IDSs and proved their effectiveness in detection intrusive behaviors [11][12].

### 3.1 Machine Learning-based IDS

The goal of any IDS is to identify two different behavior patterns: normal behavior and malicious behavior [13]. Popular machine learning algorithms that aim to cluster behavior patterns around a centroid are K-nearest neighbour and K-means [14]. As a supervised approach, Support Vector Machines (SVM) have also been be considered [15], which partitions the data plane into smaller hyperplanes and also automates feature selection.

Adaptive machine learning-based techniques for IDS have been tested in many studies to improve the accuracy of classification such as the work presented in [16] where the researchers presented a validation procedure by presenting the automated and adaptive testing prototype. the work in [17] proposed an Adaptive Model Generation (AMG) as a model generator which would work adaptively in real time to perform IDSs. AMG enables evaluation of data in real time, and automates the data collection, as well as the generation and deployment of detection models.

To the best of our knowledge, an adaptive-IDS solution for critical infrastructures that deal with both known and unknown intruders remains an open issue. In ASCH-IDS, we present a dynamic adjustment technique for the proportion of the aggregated sensory data that forwarded to the anomaly and misuse detection subsystems.

### 3.2 Deep Learning-based IDS

Deep learning methods have been applied to IDS and achieved highly accurate results [18][19][20]. In [21] the authors tested the abilities of a Deep Belief Network (DBN) in the detection of intrusive patterns. The authors in [18] combined DBN and SVMs, and introduced a hybrid IDS methodology where DBN served as a feature selector and SVM as the classifier. The hybrid approach resulted in ≈92% accuracy rate. Authors in [22] presented a partial supervised learning approach where the classifier was trained with normal traffic only so any knowledge about malicious behaviors could evolve dynamically. The authors applied Discriminative Restricted Boltzmann Machine (DRBM) to anomaly detection as an energy-based classifier.

High dimensionality is a grand challenge in big data applications; hence the authors in [23] applied an autoencoder in the first stage of an IDS in order to reduce dimensionality and extract the features for a DBN to classify anomalous and normal behavior patterns.

Deep learning has been introduced to the distributed computing such as the works presented in [24] and [25]. In [24], the authors proposed a cyber-attack detection deep learning-based mechanism in fog-to-things computing. The work in [25] proved the flexibility of cloud-based infrastructures along with the enhancements in the large-scale machine learning area for moving the highest computationally and storage tasks to the cloud to benefit of the edge computing capabilities.

### 3.3 Reinforcement learning based IDS

Reinforcement learning (RL) considered as an extension to machine learning, and involves agents that take actions to maximize the notion of rewards. Previously, reinforcement learning process was applied into IDSs by several studies. In [26], the researchers proposed a distributed reinforcement learning approach in which each agent (i.e. sensor) analyzes state observations and communicates them to a central agent. Agents that are higher in the hierarchy are equipped with the knowledge for analyzing the collected data, and they broadcast an overall abnormal state to the network operator [26]. The authors in [27] presented reinforcement learning - based technique for host-based IDS through a series of system calls by presenting a Markovian reward process (MRP) to replicate the behavior of system call series where the intrusion detection issue is transfigured to predict the MRP value functions. The work presented in [28] proposed an approach to detect intrusion with online clustering by using Pursuit Reinforcement Competitive Learning (PRCL). The researchers in [29] presented an approach of an adaptive neural networks to intrusion detection that detect new attacks autonomously with the use of reinforcement learning method. The work presented in [30] exploits the use of data flow information, Reinforcement Learning and tile coding to detect flooding-based Distributed Denial of Service (DDoS) attacks.

To this end, a comprehensive comparison/evaluation of IDS for critical monitoring infrastructures that works for both known and unknown attacks using Q-based, State-Action-Reward-State-Action Learning (SARSA) and the Temporal Difference learning (TD) reinforcement learning remains an open issue.

## 4 SYSTEM MODEL

We adopt the IDS model that was proposed for WSN-based critical infrastructure monitoring in our previous work [6] to the volunteer computing architecture.

The IDS consists of $N$ clusters with $C$ sensors in each cluster. Aggregating the sensory data is the responsibility of the Cluster Head (CH) which collects the data from sensors and directs it to the IDS which is installed in a central node namely the server. The aggregated data then undergoes the intrusion detection method. The IDS consists of : 1) Weighted cluster head selection in Section 4.1, 2) Data aggregation in Section 4.2, 3) The detection methods in Section 5. The notations used all-through the manuscript are enlisted in Table 1.

### 4.1 Weighted Cluster Head (CH) selection

A Cluster-Head (CH) election method has been adopted for data collection and processing reasons. CH election method is fulfilled by using the election technique that calculates the weight of each sensor node and compares it to the weights of the other nodes [31]. Each node has a weight $G_n$ that is a function of its ($RSSI_{sum}$) which refers to its received signal strength, mobility, and degree. Following the weight computation, the node broadcasts its weight along with its unique identifier (ID). It then proceeds to compare it with the weights of the neighbour nodes such as the sensor node that achieves the lowest weight is elected as the CH [31] [8] [6].

The weight depends on the following four elements: mobility and degree difference which refers to the node degree excluding CH capacity, cumulative time and $RSSI_{sum}$. The election procedure goes through steps as shown in Algorithm 1 Where $D_n$ refers to the degree of node $n$, $\Gamma$ represents the CH Capacity *(i.e. number of nodes)*, $\partial n$ refers to the degree difference for $n$, $SRS_{strength}(n)$ represents the sum of n's received signal strength, $\check{M}_n$ represents the mobility factor for node $n$, $\hbar_n$ refers to the cumulative time for node $n$, and $G_n$ represents the combined weight for node $n$. In Algorithm 1, $n$ refers to any node, $g_{\hbar_n}$, $g_m$, $g_{sum}$ and $g_d$ refer to the weight factors of $\hbar_n$, $\check{M}_n$, $RSSI_{sum}$ and $\partial n$, accordingly.

### 4.2 Data aggregation procedure

Each CH gathers data from its corresponding sensors and sends them to the sink. The method presented in [32] has been adopted in our previous and ongoing research as the aggregation procedure. The method depends on assessing the trust score of the aggregator by tracking the sensors' trust scores with the estimated trust between sensors and the aggregator [32]. Equation

(1) computes the trust score of a CH [32] where $T_{CH}$ denotes the CH trust value, $T_n$ is node $n$ trust value, and $T_{CH}^n$ is the CH and node $n$ trust estimation.

$$T_{CH} = \frac{(\sum_{n=0}^{n-1}(T_n + 1) \cdot T_{CH}^n)}{\sum_{n=0}^{n-1}(T_n + 1)} \quad (1)$$

---
**Algorithm 1:** Weighted Cluster Head (CH) selection pseudo-code
---
1: **procedure** WEIGHTED CH SELECTION
2: **Input:** $D_n$, $\Gamma$, $|1/SRS_{strength}(n)|$, $\check{M}_n$, $\hbar_n$
3: **Output:** $G_n$
4: **for** each node $n$ **do**
   $\partial n = |D_n - \Gamma|$
5: $|1/SRS_{strength}(n)|$:normalized value of $SRS_{strength}$
6: $G_n = g_d \partial n + \frac{g_{sum}}{|1/SRS_{strength}(n)|} + g_m \check{M}_n + g_{\hbar_n} \hbar_n$
7:    EndFor
8: **Return** $G_n$
9: **Select** the node with a minimum $G_n$ as CH
10: **Remove** CH from the nodes list
11: **Repeat** all steps for the remaining nodes
12: End
13: **end procedure**
---

## 5 INTRUSION DETECTION METHODS

### 5.1 Adaptive Machine learning based the Adaptively Supervised and Clustered Hybrid Intrusion Detection System (ASCH-IDS)

In our previously proposed ASCH-IDS [7] and CHH-IDS [6], the collected sensed data experiences two machine learning based subsystems working in parallel for intrusion detection: Anomaly Detection Subsystem ($ADSs$) and Misuse Detection Subsystem ($MDSs$). This translates into a hybrid system as shown in Algorithm 2 for detecting intrusive sensors. $ADSs$ follows the E-DBSCAN algorithm which refers to the Enhanced-Density Based Spatial Clustering of Applications with Noise while $MDSs$ follows the random forest technique. DBSCAN returns to a density based clustering method where clusters are considered as dense areas of objects where the data space is distributed into areas of objects with low densities [34] while the E-DBSCAN algorithm retains the track of the variation of local density in the cluster and computes any core objects' density variance with considering its e-neighbourhood [35]. On the other hand, when random forest used as a classification mechanism in which each tree checks each input for the frequent classes and delivers a vote if it happened [36], it operates in two stages: the training and the classification stages [37]. In [6], the collected data is partitioned into IDS subsystems (IDSs) in a round-robin fashion following a time-slotted method as shown in Fig. 1.

Fig. 2 represents the previously proposed CHH-IDS[6] where CHH-IDS runs on a WSN which contains $N$ clusters where the CH handles the aggregation procedure of the data forwarded by the sensors.

When random forest is used for misuse detection, each tree is developed as described below [37]:

- If the training set size denoted by $Y$, a randomly extracted data-points $y$ from the original data-set turn into the training data-set for the developing tree.

Table 1. Notations used in the manuscript

| Notation | Description |
|---|---|
| $V$ | Visible element |
| $H$ | Hidden element |
| $O$ | Outputs |
| $X$ | Number of visible nodes |
| $Y$ | Number of hidden nodes |
| $a_x$ | Visible bias |
| $b_y$ | Hidden bias |
| $W$ | The weights between visible and hidden layers |
| $W_{xy}$ | Combined weights of visible $V_x$ and hidden $H_y$ units |
| $\Theta$ | RBM parameters ($W_{xy}$,$a_x$,$b_y$) |
| $E(V,H|\Theta)$ | The energy function of the RBM |
| $P(V,H)$ | The probability of $(V,H)$ formation |
| $\sum_{X,Y} e^{-E(V,H)}$ | The normalization factor (all possible configurations including the visible and hidden elements) |
| $P(V)$ | The probability allocated to any visible element $V$, The network sets probability score to each case in hidden and visible elements[19][33] |
| $P(H)$ | The probability allocated to any hidden element $H$ |
| $P(V|H)$ | The probability of $V$ independently with $H$ |
| $P(H|V)$ | The probability of $H$ independently with $V$ |
| $P(Intrusive)$ | The binary output for intrusive detection |
| $Normal$ | The binary output for normal detection |
| $rec$ | Number of records |
| $Var$ | Number of variables |
| $T_r$ | Number of trees |
| $TP$ | True Positive |
| $TN$ | True Negative |
| $FP$ | False Positive |
| $FN$ | False Negative |
| $S_t$ | State at time $t$ |
| $A_t$ | Action at time $t$ |
| $R_t$ | Reward at time $t$ |
| $V^\pi(S)$ | Value estimation of $R$ that at initial state $S$ |
| $\pi(S,A)$ | Probability of $A$ in $S$ |
| $P_{SS^+}(A)$ | Transitional probability from state $S$ to $S^+$ at $A$ |
| $R(S,S^+,A)$ | Reward returned from transition from state $S$ to $S^+$ at $A$ |
| $r$ | Discount factor weight from future rewards to current rewards |
| $V^\pi_I(S^+)$ | Value estimation of $R$ at state $S^+$ at the initial iteration $I$ |
| $V^\pi_{I+1}(S)$ | Value estimation of $R$ at state $S$ at the updated iteration $I+1$ |
| $T_{CH}$ | CH trust value |
| $T_n$ | Node $n$ trust value |
| $T^n_{CH}$ | CH and node $n$ trust estimation |
| $\alpha$ | The weight of the recently calculated $TP/FP$ and its value during the ($\triangle \tau$) |

- If the input variables denoted by $X$, a randomly extracted data-points $x$ from $X$ can be utilized to split the node. $x$ considered as a constant while the forest develop where each tree is developed to the most size.

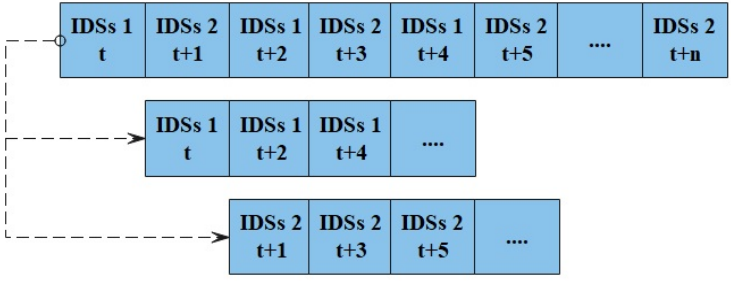

Fig. 1. Aggregated sensed data distribution mechanism between the IDSs1 and IDSs2.

On the other hand, E-DBSCAN used as clustering algorithm which has been utilized to extract the $\epsilon$ which represents the distance threshold, $\epsilon$ is a critical factor in E-DBSCAN algorithm [38]. DBSCAN can realize the adjacent clusters which belong to the same density as well as clusters of random shapes [39]. DBSCAN consists of two factors, the $\epsilon$ and the $MinPts$ which considered as input parameters. It follows the rules descried below:

- $N_\epsilon(x) = y \in X|d(x,y)| \leq \epsilon$ is the $\epsilon - neighbourhood$ of point $x$.
- Neighbourhoods of core object has a size $> MinPts$.
- A point $j$ is density accessible from $i$ as a core object.
- $i$ and $j$ considered as density-based connected when $i$ and $j$ are density accessible from a core object.

On the other hand, the ASCH-IDS [6] retains the track of $ADSs$ and $MDSs$ deviations in the Receiver Operating Characteristics (ROC) and changes the segment of data directed to them adaptively.

Equations. (2)-(3) represent the $(TP)$ to $(FP)$ for the two subsystems at $\tau_i$ which are represented by $\mu_1(\tau_i)$ and $\mu_2(\tau_i)$ [7].

---

**Algorithm 2:** Detection of intrusive sensors

1: **procedure** INTRUSIVE SENSORS DETECTION
2: **Cluster Head (CH) selection:** Weighted CH selection
3: **Data aggregation:** Trust-based aggregation of sensed data
4: **Data distribution:** Aggregated data distribution between the two sub-systems
5: **if** *(Intrusive behavior detected)* **then**
       ALARM
6: END
8: **else**
       **if** *(End of sensing data)* **then**
       END
10: **else**
       Go to **Data distribution**
12: **end procedure**

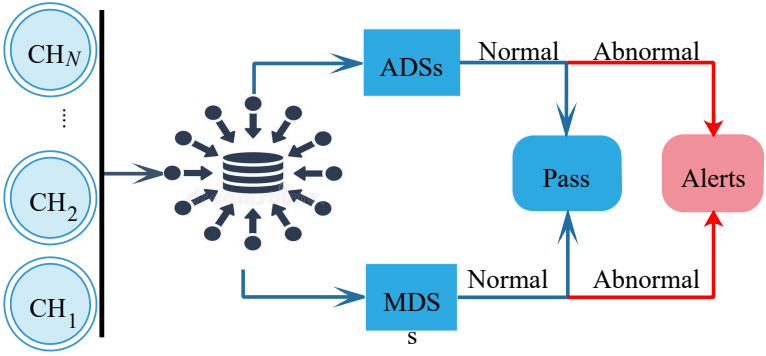

Fig. 2. CHH-IDS system model. The collected sensed data is distributed between the ADSs and MDSs.

$$\mu_1(\tau_i) = \frac{TP_1(\tau_i)}{FP_1(\tau_i)} \quad (2)$$

$$\mu_2(\tau_i) = \frac{TP_2(\tau_i)}{FP_2(\tau_i)} \quad (3)$$

Equations (4)-(5) represent the $TP$ to $FP$ ratios at time $(\triangle\tau)$ where $\triangle\tau=\tau_{i+1} - \tau_i$ [7].

$$\mu_1(\triangle\tau) = \frac{TP_1(\triangle\tau)}{FP_1(\triangle\tau)} \quad (4)$$

$$\mu_2(\triangle\tau) = \frac{TP_2(\triangle\tau)}{FP_2(\triangle\tau)} \quad (5)$$

When the ROC behavior during $\triangle\tau$ is extracted, the ROC behavior of each subsystem can be obtained as the sum of the ROC behavior at $\tau$ and the behavior at $\triangle\tau$ as represented in Eqs. (6) and (7) [7]. where $\alpha$ represents the weight of the recently calculated $TP/FP$ and its value during the $(\triangle\tau)$ where $\tau_{i+1} = \tau_i + \triangle\tau$.

$$\mu_1(\tau_{i+1}) = \alpha\mu_1(\tau_i) + (1 - \alpha)\mu_1(\triangle\tau) \quad (6)$$

$$\mu_2(\tau_{i+1}) = \alpha\mu_2(\tau_i) + (1 - \alpha)\mu_2(\triangle\tau) \quad (7)$$

Eq.(8) represents an indicator $I(\tau_i)$ which tracks the ROC behavior of MDSs and ADSs at any time $\tau_i$.

$$I(\tau_i) = \frac{\mu_1(\tau_i)}{\mu_2(\tau_i)} \quad (8)$$

The extracted indicator $I$ is used in the decision of directing the collected data such as if $I(\tau_i) > I(\tau_{i-1})$, ASCH-IDS decides that *ADSs* is performing better than *MDSs*. so, any increasing of data proportion on *ADSs* is enhancing the overall performance of the system. variously, if $I(\tau_i) < I(\tau_{i-1})$, the intrusion system determines that *MDSs* is performing better than *ADSs*, such as, any increase in the data proportion on *MDSs* is expected to improvement of the overall performance of the system. wherever, if $I(\tau_{i+1}) > I(\tau_i)$, the ASCH-IDS increases the data proportion on $\mu_1$ and decreases on $\mu_2$ such as: $D_a(\tau_{i+1}) = D_a(\tau_i) + \triangle D$ and $D_m(\tau_{i+1}) = D_m(\tau_i) - \triangle D$ as formulated in (9)-(10) where $\triangle D$ denotes the data adjustment for each subsystem. Algorithm 3 presents an overview of the previous steps. Where $\tau_i, \tau_{i+1}$ and $\triangle\tau$ refer to any time, $\tau_i + \triangle\tau$ and the time difference between $(\tau_{i+1} - \tau_i)$ respectively. $\alpha$ refers to the ROC characteristics weight in the evaluation of $\mu_1(\tau_i)$ and $\mu_2(\tau_i)$, $D_a(\tau_i)$ and $D_m(\tau_i)$ refer to the segments of data forwarded to the anomaly (ADSs) and the misuse (MDSs) at $\tau_i$.

$$D_a(\tau_{i+1}) = D_a(\tau_i) \pm \triangle D \tag{9}$$

$$D_m(\tau_{i+1}) = D_m(\tau_i) \pm \triangle D \tag{10}$$

As the proposed system consists of two subsystems, its complexity is a function of the two algorithms complexities. Since the random Forest is a special model of decision trees, its complexity can be extracted from decision tree, such as the complexity $C(Random forest)$ for building a decision tree with $r$ records and $v$ variables is $O(v + rec * \log(v))$, while for multi trees as in our random forest, the complexity is $O(Tr * Var * rec * log(Var))$ where $T_r$ is the number of trees, $Var$ is the number of variables. For E-DBSCAN as the second subsystem, the complexity is directed by the query requests' number. Since each point is operates with only one query, the run-time complexity is $O(n) < (n * \log(n))$. In our E-DBSCAN, the initialization step is executed 1 time, the comparison step is executed $(m+1)$ times, and the incremental step is executed $(m)$ times. Thus, the complexity will be as $O(2 + (2m)) = O(m)$. To this end, the overall complexity of the system can be calculated by $O((T_r * Var * rec * \log(Var)) + (2 + (2m))$.

## 5.2 Deep learning based Restricted Boltzmann Machine-based Clustered IDS (RBC-IDS)

Restricted Boltzmann machine (RBM) is a neural, energetic network containing two types of layers: (V) and (H) which refer to the visible and the hidden layers respectively, where the learning procedure is steered by an unsupervised fashion [19]. The RBM permits connections between neurons of the same layer, which makes it restricted, and the procedure is presented in the pseudo-code in Algorithm 4.

Table 1 contains the RBM parameters used in Algorithm 4.

The network sets probability score to each case in hidden and visible elements[19][33].

Fig. 3 represents the used RBM setting in RBC-IDS. The RBC-IDS consists of an input layer contains $X$ visible nodes, three hidden layers and output layer with two outputs $O_1$ and $O_2$ for *Intrusive* and *Normal* outputs respectively. $W_{11}$ represents the weight between the first visible layer and the first hidden layer while $W_{12}$ refers to the weight of the first and the second hidden layers and $W_{23}$ is the weight between the second and the third hidden layers.

In the RBC-IDS, each CH is responsible for aggregating the sensed data from the sensors in the same cluster and forwards them to the server by adopting the procedure in [32].

## 5.3 Reinforcement learning

In machine learning, the studied environment is defined as Markov Decision Process (MDP) which as many reinforcement learning algorithms such as Q-learning, employ dynamic programming procedures. MDP come up with optimum policy to achieve the maximum rewards over time [40]. The fundamentals of reinforcement learning concept are as follows:

- The agent interacts with the environment and takes actions $A_t$ in each state $S_t$ and observes the feedback from the environment.
- The environment supplies a reward $R_t$ for the actions performed in form of $R^+$ or $R^-$ which refer to positive or negative rewards respectively.
- The agent observes the environment for any changes and optimize the received rewards by updating the policies.
- Different reinforcement learning techniques are called in order to maximize the expected value of the total reward, starting from the current state.

*5.3.1 Q-Learning.* Q Learning builds on the concept of value iteration where the agent aims to estimate the value function to update all states *s* and actions *A* every iteration in order to know

---

**Algorithm 3:** Adaptive IDS: (ASCH-IDS)
---

1: **procedure** ROC TRACKING ▷ Tracking the deviations in (ROC) of the anomaly and the misuse detection subsystems
2: **Input:** $TP_1(\tau_0)$, $FP_1(\tau_0)$, $TP_2(\tau_0)$, $FP_2(\tau_0)$
3: **Output:** $\mu_1(\tau_i)$, $\mu_2(\tau_i)$
4:          ▷ $\mu_1(\tau_i)$: Ratio of $TP$ to $FP$ at $\tau_i$ for the anomaly detection subsystem
5:          ▷ $\mu_2(\tau_i)$: Ratio of $TP$ to $FP$ at $\tau_i$ for the misuse detection subsystem
6:
7: **Initiate:** $\mu_1(\tau_0) \leftarrow$ INIT, $\mu_2(\tau_0) \leftarrow$ INIT, $t_i \leftarrow \tau_0$, $I(\tau_0) \leftarrow 1$   ▷ $\mu_1(\tau_0) = TP_1(\tau_0)/FP_1(\tau_0)$
8:                       ▷ $\mu_2(\tau_0) = TP_2(\tau_0)/FP_2(\tau_0)$
9: **for** *Allnetwork* **do**
  **if** *HALT* **then**
   END
  **else**
   Check for *NextAggregation*
10:   **if** *NextAggregation* **then**
13:    $\tau_i \leftarrow \tau_i + \Delta\tau$
15:
16:    $\mu_1(\tau_i) \leftarrow \alpha\mu_1(\tau_i - \Delta\tau) + (1-\alpha)\frac{TP_1(\tau_i) - TP_1(\tau_i - \Delta\tau)}{FP_1(\tau_i) - FP_1(\tau_i - \Delta\tau)}$
17:
18:    $\mu_2(\tau_i) \leftarrow \alpha\mu_2(\tau_i - \Delta\tau) + (1-\alpha)\frac{TP_2(\tau_i) - TP_2(\tau_i - \Delta\tau)}{FP_2(\tau_i) - FP_2(\tau_i - \Delta\tau)}$
19: **else**
   Go to *HALT*
20:   **if** $I_{(\tau_i)} > I_{(\tau_i - \Delta\tau)}$ **then**
23:    $D_a(\tau_i) \leftarrow D_a(\tau_i - \Delta\tau) + \Delta D$
24:
25:    $D_m(\tau_i) \leftarrow D_m(\tau_i - \Delta\tau) - \Delta D$
26:
27:    Go to Initiate
28:
  **if** $I_{(\tau_i)} < I_{(\tau_i - \Delta\tau)}$ **then**
30:    $D_a(\tau_i) \leftarrow D_a(\tau_i - \Delta\tau) - \Delta D$
31:
32:    $D_m(\tau_i) \leftarrow D_m(\tau_i - \Delta\tau) + \Delta D$
33:
34:   Go to Initiate **else**
   Go to Initiate
36:
37:  **END**
38: **end procedure**

---

which action $A$ leads to higher reward $R$. In the Q table, the rows represent the states while the columns represent the actions. In a state (say state $S$), the agent takes an action (i.e. action $A$), observes the reward ($R$) for this action, as well as the next state ($S'$), and re-estimates the Q value.

Equation (11) represents the estimated $Q$ value [41] where $S_t$, $A_t$ and $R_t$ stand for the state, action and reward at time $t$, respectively. In addition, $\alpha$ and $\gamma$ represent the learning rate and a constant regarding the relative value of rewards, respectively. Indeed, the following conditions hold for both parameters: $0 < \alpha < 1$ and $0 < \gamma < 1$.

$$Q^+(S_t, A_t) \leftarrow (1-\alpha)Q^+(S_t, A_t) + \alpha(R_t + \gamma max Q^+(S_t^+, A_t^+)) \tag{11}$$

Equation (12) formulates the estimated function $V^\pi(S)$ which represents the estimation of the future reward $R$ that will be received in the initial state $S$ [42]. In the equation $\pi(S, A)$ denotes the likelihood of action $A$ in state $S$, $P_{SS^+}(A)$ stands for the state transition probability between states $S$ and $S^+$ with action $A$. $R(S, S^+, A)$ is the reward issued after transitioning from state $S$ to $S^+$ at action $A$, $r$ is the discount factor weight from future rewards to current rewards[42]. The value iteration used method is shown in Eq.(13) below. Where $V^\pi(S)$ refers to the value estimation of $R$ that at initial state $S$, $\pi(S, A)$ is the probability of $A$ in $S$, $P_{SS^+}(A)$ refers to the transitional probability from state $S$ to $S^+$ at $A$, $R(S, S^+, A)$ is the reward returned from transition from state $S$ to $S^+$ at $A$, $r$ is the discount factor weight from future rewards to current rewards, $V_I^\pi(S^+)$ is the value estimation of $R$ at state $S^+$ at the initial iteration $I$, and $V_{I+1}^\pi(S)$ is the value estimation of $R$ at state $S$ at the updated iteration $I + 1$.

---

**Algorithm 4:** Restricted Boltzmann Machine (RBM) Procedure

---

1: **procedure** RBC-IDS
2: **Input:** $W_{xy}$, $a_x$, $b_y$
3: **Output:** $P(Intrusive), P(Normal)$
4: **Initiate:** $W_{xy}$, $a_x$, $b_y$, $X$, $Y$.
     **for** $y = 1, 2, 3, ..., Y$ *(All hidden layers)* **do**
     **for** $x = 1, 2, 3, ..., X$ *(All visible layers)* **do**
        **Compute** $E(V, H|\Theta) = -\sum_{x=1}^{X} a_x V_x - \sum_{y=1}^{Y} b_y H_y - \sum_{x=1}^{X} \sum_{y=1}^{Y} V_x H_y W_{xy}$
        **Compute** $P(V, H) = \frac{e^{-E(V,H)}}{\sum_{X,Y} e^{-E(V,H)}}$
5:
8: **Compute** $P(V) = \sum_Y P(V, H) = \frac{\sum_Y e^{-E(V,H)}}{\sum_X \sum_Y e^{-E(V,H)}}$
9:
10: **Compute** $P(H) = \sum_X P(V, H) = \frac{\sum_X e^{-E(V,H)}}{\sum_X \sum_Y e^{-E(V,H)}}$
11:
12: **Compute** $P(V|H) = \prod_{x=1}^{V} P(V_x|H)$  **Compute** $P(H|V) = \prod_{y=1}^{H} P(H_y|V)$
13:
14: **Compute** $P(Intrusive) = P(O = 0|V)$
15: **Compute** $P(Normal) = P(O = 1|V)$
16: End for
17: End for
18:
19:
20: **end procedure**

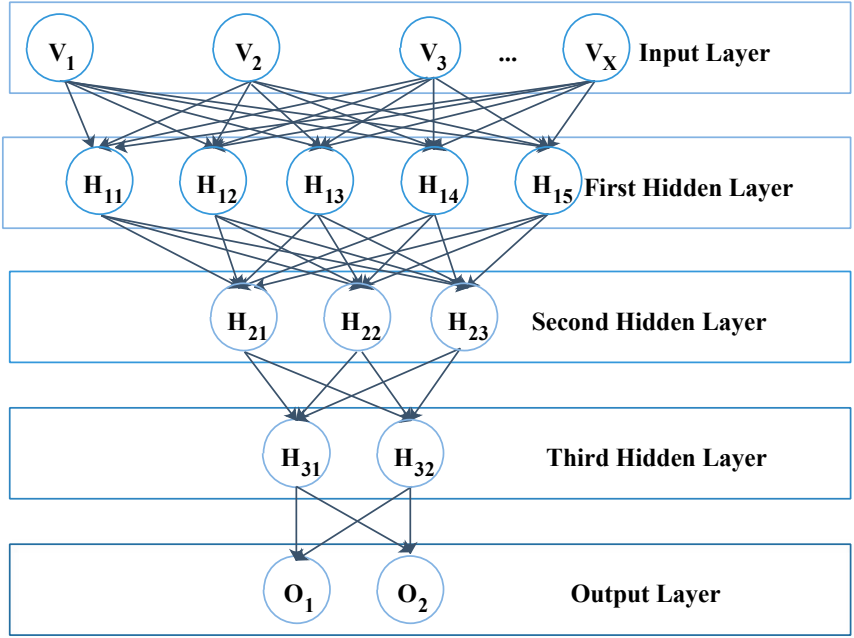

Fig. 3. RBM used in RBC-IDS which contains one visible layer and 3 hidden layers and one output layer. $W_{11}$ refers to the weight between the visible layer and the first hidden layer, $W_{12}$ refers to the weight between the first and second hidden layers and $W_{23}$ refers to the weight between the second and third hidden layers. $O_1$ and $O_2$ are the *Intrusive* and *Normal* outputs.

$$V^\pi(S) = \sum_A \pi(S, A) \sum_{S^+} P_{SS^+}(A) R(S, S^+, A) + rV^\pi(S^+) \tag{12}$$

$$V^\pi_{I+1}(S) = max_A \sum_{S^+} P_{SS^+}(A) R(S, S^+, A) + rV^\pi_I(S^+) \tag{13}$$

It is worth noting that the reason behind adopting Q-learning as one of our reinforcement learning methods is the model-free nature of Q-learning. Furthermore, by applying Q-learning, it is also possible to address stochastic rewards in a non-adaptive manner. In addition, Q-learning has the ability to learn without necessarily following the current policy [43].

*5.3.2 State-Action-Reward-State-Action Learning (SARSA).* SARSA is an MDP-based reinforcement learning algorithm which is considered as a Modified Connectionist Q-Learning (MCQ-L) algorithm. SARSA updates the Q-values depending on current state $S$, the current action $A$ for $S$, the returned reward $R$ of action $A$, the new state $S$, and the next action $A$ for the new state, which can be represented in the quintuple $(S_t, A_t, R_t, S_{t+1}, A_{t+1})$.

SARSA in an on-policy learning algorithm where the agent interrelates with environment and updates the policy considering the taken actions. In SARSA, the Q-value function represents the received reward in the next time step for taking action $A$ in state $S$, and the reward received from the next state and action [44]. The previous Q-value function (Eq.(11)) can be updated as in Eq.(14).

The equations Eq.(11) and Eq.(14) look almost the same except that in Q-learning the action with highest estimation of all possible next actions will be considered while the actual next action is considered in SARSA. Looking for the maximum reward in Q-learning can make it more costly compared to SARSA technique [14].

$$Q(S_t, A_t) \leftarrow Q(S_t, A_t) + \alpha(R_t + \gamma Q^+(S_{t+1}, A_{t+1}) - Q(S_t, A_t)) \tag{14}$$

*5.3.3 Temporal Difference learning (TD) .* is a model-free reinforcement learning technique which can learn by estimating the expected value function by considering approximating distribution from the current value. The TD technique estimates the state value function under a policy $\pi$ as shown in Eq.(15) and Eq.(16) bellow.

$$V^\pi(S_t) = E_\pi\{\sum_{t=0}^{\infty} \gamma^t R_t\} \tag{15}$$

$$V^\pi(S_t) = E_\pi\{R_0 + \gamma V^\pi(S_1)\} \tag{16}$$

Where $V^\pi(S_t)$ refers to the state value function with state $S_t$, $R$ refers to the reward and $\gamma$ is the discount rate under the policy $\pi$. In Eq.(16), $R_0 + \gamma V^\pi(S_1)$ represents the unbiased estimate for $V^\pi(S_t)$. TD is a technique used to learn how to estimate a value that depends on future values, which make it useful to learn both the V- function and the Q- function. Whereas, Q-learning is a specific technique to learn only the Q-function.

Reinforcement learning-based Intrusion Detection System (QL-IDS) for WSNs is illustrated in Fig. 4. The IDS consists of hierarchically connected clusters with the aggregator and one central agent its the Agent in the representation of Reinforcement learning Box which tries to model the state of the monitored network. Following a series of iterations, the central agent knows the action $A$ that need to be executed in response to each state $S$ in order to obtain a positive reward $R^+$.

It is worth mentioning that the value iteration works by generating consecutive estimates of the optimal value function. I n which each of these iterations can be accomplished in $O(|A||S|^2)$. In reinforcement learning the required number of iterations can grow exponentially.

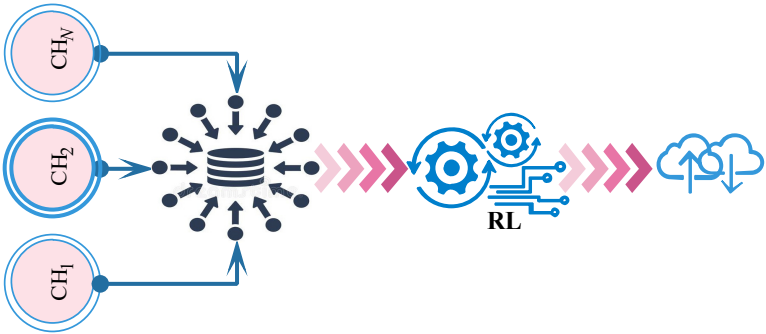

Fig. 4. Reinforcement Learning based IDS representation

## 6 EXPERIMENTAL RESULTS AND ANALYSIS

### 6.1 Description of the KDD data-set

We simulate 20 devices in a network that communicate through H-DSR protocol which is a Dynamic Source Routing protocol specialized for the Hierarchical networks [45]. The tested devices are deployed in four clusters in an area of 100m x 100m. Resulting figures present the mean of 10 tests (runs) for each scenario with confidence level of 95%. Table 2 presents the simulation inputs.

Table 2. Testing settings

| Simulation Inputs | Input Value |
|---|---|
| Visible elements (inputs) | 41 |
| Hidden layers | 3 |
| Number of sensors | 20 |
| Operational area | 100m x 100m |
| Number of clusters | 4 |
| Simulation time | 600s |
| Attack Types | DoS,Probe,U2R,R2L |
| Communication range | 100m |
| $\alpha$ (The weight of $TP/FP$) | 0.7 |
| $INIT$ | 0.5 |
| Routing protocol | H-DSR |
| Packet size | 250 bytes |
| Trust range | [0,1] |

The MIT Lincoln laboratory has collected the data-sets for computer network IDS evaluation under the support of Defense Advanced Research Projects Agency (DARPA) and Air Force Research Laboratory (AFRL). The Knowledge Discovery in Data mining CUP 1999 (KDDCup99) data-set is a subset of the DARPA data-set [46]. The KDDCup99 is used to test the IDSs efficiency on the simulated WSN, with each connection record containing 41 features and which are classified as normal or attack behaviors. The KDDCup99 contains around 311,029 records as test data-set and 494,020 records as training data-set. The numerous attack types in these data-sets are gathered into attack groups that assign similar attack types within a single group, resulting in an advancement of the detection rate [46]. KDDCup99 attacks fall under four major groups, specifically, DoS, Probe, R2L and, U2R, which refer to Denial of Service, probe, Remote to Local and, User to Root attacks respectively. Table 3 presents examples of the attacks in the KDDCup99 test data-set and KDDCup99 training data-set.

Table 3. KDD data set description [47]

| Attack | Attacks in test set | attacks in training set |
|---|---|---|
| Dos | processtable, mailbomb | neptune, teardrop |
| R2L | named, xsnoop | spy, multihop |
| U2R | ps, xterm | bufferoverflow, perl |
| Probe | saint, mscan | portsweep, ipsweep |

## 6.2 Performance Evaluation

IDSs are evaluated based on the following criteria's.
(1) **True Positive (TP)**: Are the anomalous cases that were correctly classified as abnormal.
(2) **False Positive (FP)**: Are the normal cases that were incorrectly classified as anomalous.
(3) **True Negative (TN)**: Are the normal cases that were classified correctly.
(4) **False Negative (FN)**: Are the anomalous cases that were incorrectly classified as normal.

*6.2.1 Pre-processing Phase.* Since the models performance evaluation are tested using KDD'99 dataset, some of the features are represented by string values (i.e protocols names),*Numerical*

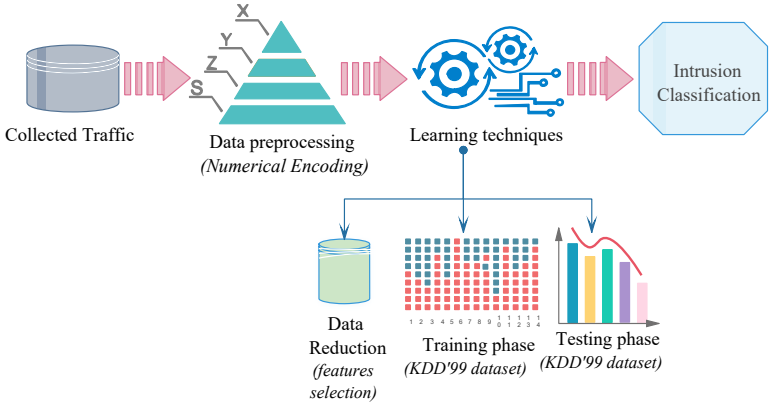

Fig. 5. The adopted learning phases

*encoding* process has been adopted for this matter. The TCP, ICMP and UDP protocols names were encoded as 001, 010 and 011, respectively.

*6.2.2 Training and Testing Phase.* The used KDD'99 dataset consists of training and testing datasets. Each data line of the dataset has 41 features, consist of 38 numerical features and three non-numerical features. In NS3 we start with training the different models (ex. RBC-IDS training started with training the first layer, gathering the data generated from the trained layer and using these data as the training dataset for the second layer and so on).

Fig. 5 represents the adopted common phases between all the proposed learning mechanisms (ex. machine learning, deep learning and reinforcement learning).

*6.2.3 Evaluated Metrics.* The following metrics have been considered for evaluating the various IDSs.

**Accuracy Rate (AR)**: AR denotes the ratio of the truly classified incidences that return to True Positive (TP) and True Negative (TN) incidences. AR is represented in Eq. (17) [48].

$$AR = \frac{TP + TN}{TP + TN + FP + FN} \quad (17)$$

*AR* has been presented to trace ASCH-IDS, RBC-IDS, TD-IDS, SARSA-IDS and the QL-IDS ARs for different scenarios. Fig. 6 shows the *AR* for the ASCH-IDS, RBC-IDS, TD-IDS, SARSA-IDS and the QL-IDS. The proposed QL-IDS achieves the highest *AR* of 100% followed by SARSA-IDS with *AR* of 99.97% and TD-IDS with 99.94%.

QL-IDS performs with the highest *AR* for two reasons: First, Q-Learning is based on the exemplars from the training data-sets and suited for decision making while the system is running. The agent interacts with the environment and aims to optimize the cumulative reward ($R^+$) by learning the best actions through feedback. On the other hand, RBC-IDS makes predictions about data and learns from training data-sets to build a classifying model.

**Detection Rate (DR)**: DR denotes the behavior that is accurately recognized as intrusive. It signifies the (*TP*) as displayed in Eq. (18)[48]. *DR* for ASCH-IDS, RBC-IDS, TD-IDS, SARSA-IDS and QL-IDS are shown in Fig.(7).

$$DR = \frac{TP}{TP + FP} \quad (18)$$

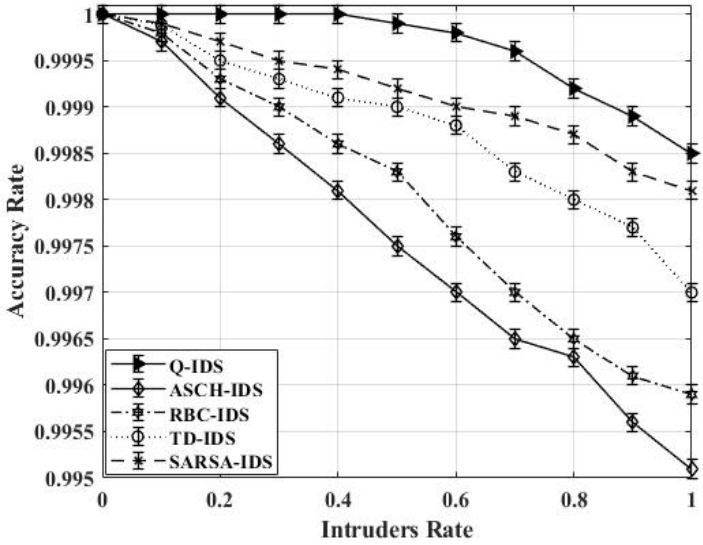

Fig. 6. Accuracy rate comparison between QL-IDS,ASCH-IDS, RBC-IDS, TD-IDS and SARSA-IDS.

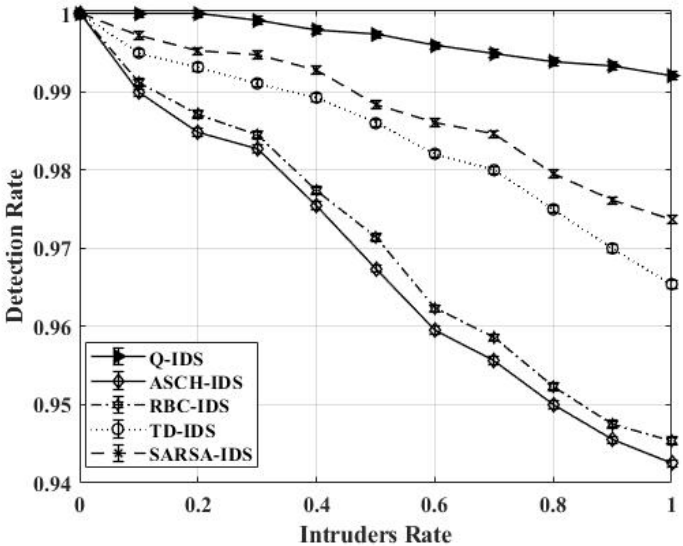

Fig. 7. Detection rates comparison between QL-IDS,ASCH-IDS,RBC-IDS, TD-IDS and SARSA-IDS

Fig. 7 illustrates the *DRs* for ASCH-IDS, RBC-IDS, TD-IDS, SARSA-IDS and QL-IDS. As shown in the figure, the proposed QL-IDS achieves the highest *DR* followed by SARSA-IDS and TD-IDS.

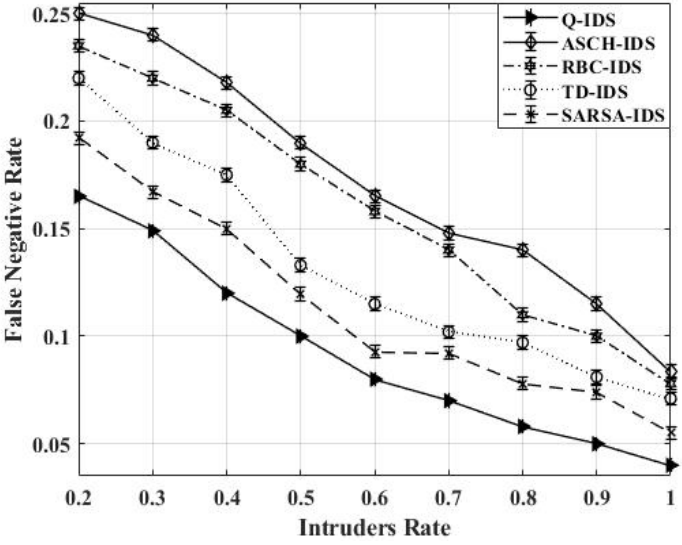

Fig. 8. FNR comparison between QL-IDS, ASCH-IDS, RBC-IDS, TD-IDS and SARSA-IDS

**False Negative Rate (FNR)**: FNR defined as the ratio of malicious behavior that have been designated as non-intrusive[6], as shown in 19 [48].

$$FNR = \frac{FN}{TP + FN + FP + TN} \qquad (19)$$

FNR of ASCH-IDS, RBC-IDS, SARSA-IDS, TD-IDS compared to QL-IDS is shown in Fig. 8.

In reinforcement learning-based QL-IDS, the overall FNR has been mitigated when compared to the case under the other reinforcement learning techniques (SARSA-IDS and TD-IDS), deep learning-based RBC-IDS and the ASCH-IDS reacting to the increase of $TP$ which is the reason behind the $DR$ and $AR$ enhancement.

This can be interpreted as follows: While QL uses function approximation that represents the value function to react to all actions and target the positive rewards, RBM is competent at feature reduction which helps in the elimination of redundant features and reduction of the FN cases.

**(ROC) curve**: The ROC curve refers to the ratio between Sensitivity and the $(1 - Specificity)$ which are $(TP)$ and $(FP)$ respectively.

**Sensitivity-specificity** ratio is represented by the area under the curve where the larger area reflects the best performance. We plot the ROC curves for ASCH-IDS, QL-IDS, RBC-IDS SARSA-IDS and TD-IDS in order to assess the system performance as shown in Fig. 9. It is clear that the QL-IDS performs better with the largest area under the curve followed by SARSA-IDS and TD-IDS. Since QL-IDS achieves the highest $TP$, the QL-IDS achieves the best performance in term of ROC.

$F_1$ **Score** : $F_1$ score studies the **precision-recall** of the test in order to calculate its $F_1$ score. The **precision** is the number of true positive incidences divided by all positive incidences, it is expressed as $TP/(TP + FP)$. The **recall** is formulated as $TP/(TP + FN)$, representing the number of true positive incidences divided by all *actually* positive instances as shown in Fig. 10. High system performance can be achieved by the high recall and high precision. Thus, the near precision-recall to 1, results in the best performance [49]. QL-IDS achieves the highest precision-recall ratio compared to ASCH-IDS and RBC-IDS as shown in figure 10. Precision and recall primarily depend on TP

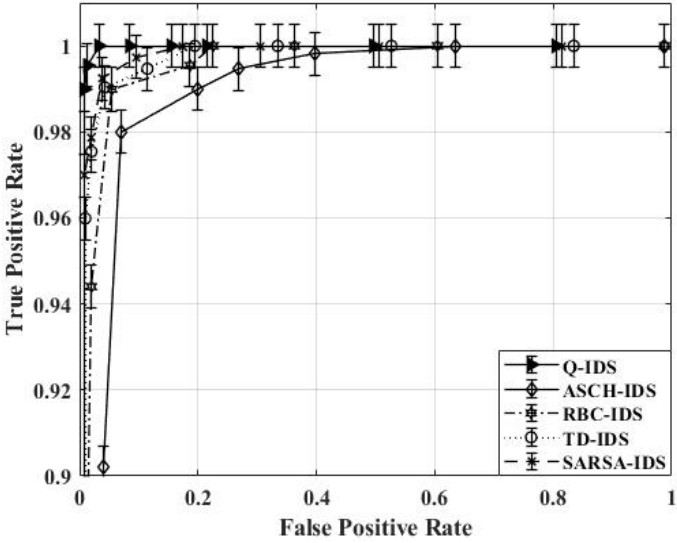

Fig. 9. ROC curves representation comparison between QL-IDS,ASCH-IDS,RBC-IDS, SARSA-IDS and TD-IDS

!htbp

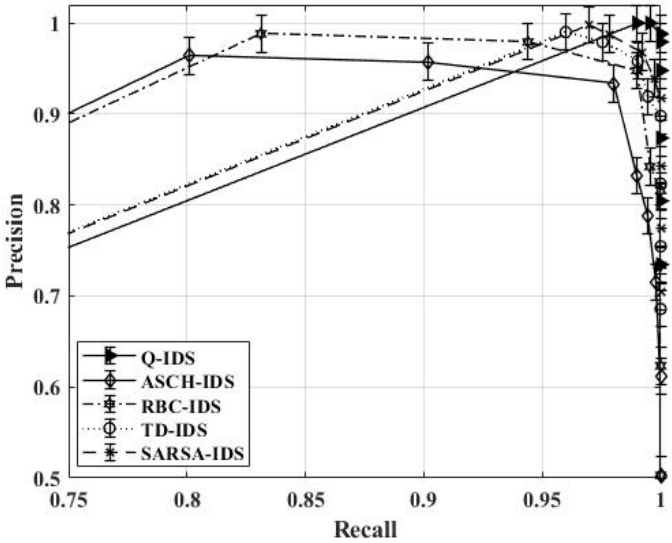

Fig. 10. Precision-Recall representation of QL-IDS, ASCH-IDS, RBC-IDS, SARSA-IDS and TD-IDS

performance. QL-IDS performs with the highest *precision* − *recall* since Q-learning is based on the exemplars from the training data-sets and suited for on-the-fly decision making while the agent interacts with the environment tested environment and uses the feedback to choose the actions to optimize the cumulative reward ($R^+$).

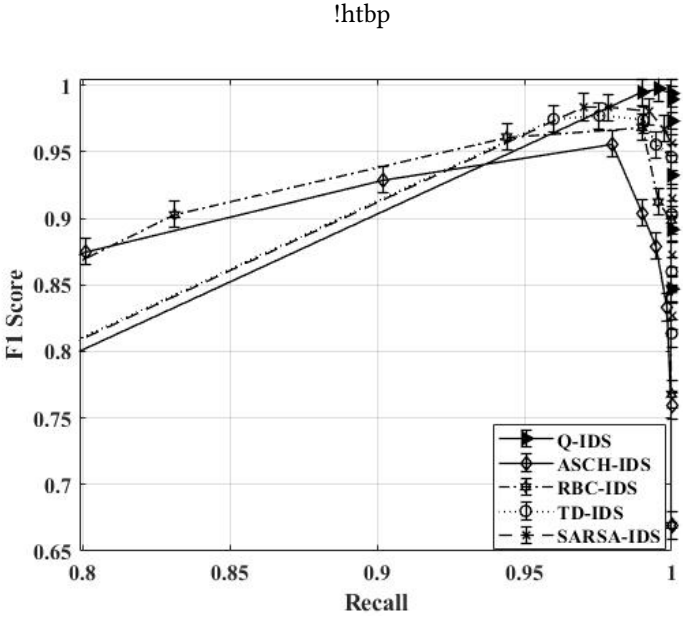

Fig. 11. $F_1$ Score representation of QL-IDS, ASCH-IDS, RBC-IDS, SARSA-IDS and TD-IDS

$F_1$ denotes the harmonic average of precision and recall, $F_1$-score is formulated as in eq. (20), and $F_1$-score performance of the IDS solutions for the WSNs under investigation are plotted in Fig. 11 [50].

$$F_1 = 2 * \frac{Precision * Recall}{Precision + Recall} \qquad (20)$$

The $F_1$ score measures test accuracy [50] by studying the precision-recall of the test in order to calculate its $F_1$ score. $F_1$ score is a reliable metric to use if the test cases seek balance between precision and recall. Thus, the better the precision-recall the better the $F_1$ score. Figure 11 shows that QL-IDS introduces improved $F_1$ score compared to SARSA-IDS, TD-IDS, ASCH-IDS and RBC-IDS.

## 7 CONCLUSION

Volunteer computing uses Internet-connected devices make it easy for participants to share their resources especially for the critical infrastructures. Critical infrastructure security is considered as one of the vital issues in smart cities. Monitoring the networks components and detecting malicious behaviors are fundamental functions to ensure security of the monitoring operation. Since Wireless Networks are deployed in open and uncontrolled environments, transmitting data through them leads to huge vulnerabilities. Therefore, the robustness of Intrusion Detection Systems (IDSs) in Wireless networks is a must. In this study, we present a comparative study between machine learning based IDS systems with deep learning based IDS for critical infrastructure monitoring WSNs. We specifically investigate our previously proposed machine learning-based Adaptively Supervised and Clustered Hybrid IDS (ASCH-IDS) with a Restricted Boltzmann Machine-based and Clustered IDS (RBC-IDS), Q-Learning-based IDS (QL-IDS), SARSA-IDS and TD-IDS, which are reinforcement learning solutions. Through simulations, we have verified that QL-IDS works with ≈ 100% detection rate and ≈ 100% accuracy rate in the existence of intrusive behavior in the tested

WSN. From the performance analysis, we have shown that the adaptive machine learning-based solution performs in same rates as the deep learning-based solution whereas adopting a machine learning-based IDS framework leads to approximately half the detection time of the deep learning-based RBM-IDS framework. We have also shown that the reinforcement learning-based solutions perform with the best precision-recall, $F_1$ score with $\approx 1$ which represents the best performance and with the largest area under curve (ROC).

As a future work, we are planning to extend the presented IDS to larger networks that consist of large number of nodes and clusters. Moreover, we are testing the impact of heterogeneous cluster sizes on our presented solution's overall performance.

## ACKNOWLEDGMENTS

This work was supported in part by the Natural Sciences and Engineering Research Council of Canada (NSERC) DISCOVERY Program.

A Comparative Study of AI-based Intrusion Detection Techniques in Critical Infrastructures       21[18] Mostafa A. Salama, Heba F. Eid, Rabie A. Ramadan, Ashraf Darwish, and Aboul Ella Hassanien. Hybrid intelligent intrusion detection scheme. In António Gaspar-Cunha, Ricardo Takahashi, Gerald Schaefer, and Lino Costa, editors, *Soft Computing in Industrial Applications*, pages 293–303, Berlin, Heidelberg, 2011. Springer Berlin Heidelberg.

[19] Arnaldo Gouveia and Miguel Correia. *A Systematic Approach for the Application of Restricted Boltzmann Machines in Network Intrusion Detection*, volume 10305. 05 2017.

[20] Yazan Otoum, Dandan Liu, and Amiya Nayak. Dl-ids: a deep learning–based intrusion detection framework for securing iot. *Transactions on Emerging Telecommunications Technologies*, n/a(n/a):e3803. e3803 ett.3803.

[21] M. Z. Alom, V. Bontupalli, and T. M. Taha. Intrusion detection using deep belief networks. In *National Aerospace and Electronics Conference (NAECON)*, pages 339–344, June 2015.

[22] Ugo Fiore, Francesco Palmieri, Aniello Castiglione, and Alfredo De Santis. Network anomaly detection with the restricted boltzmann machine. *Neurocomputing*, 122:13 – 23, 2013.

[23] Yuancheng Li, Rong Ma, and Runhai Jiao. A hybrid malicious code detection method based on deep learning. 9:205–216, 05 2015.

[24] A. Abeshu and N. Chilamkurti. Deep learning: The frontier for distributed attack detection in fog-to-things computing. *IEEE Communications Magazine*, 56(2):169–175, Feb 2018.

[25] Rafał Kozik, Michał Choraś, Massimo Ficco, and Francesco Palmieri. A scalable distributed machine learning approach for attack detection in edge computing environments. *Journal of Parallel and Distributed Computing*, 119:18 – 26, 2018.

[26] Arturo Servin and Daniel Kudenko. Multi-agent reinforcement learning for intrusion detection. In *Adaptive Agents and Multi-Agent Systems III. Adaptation and Multi-Agent Learning*, pages 211–223, Berlin, Heidelberg, 2008. Springer Berlin Heidelberg.

[27] Xin Xu and Tao Xie. A reinforcement learning approach for host-based intrusion detection using sequences of system calls. In *Advances in Intelligent Computing*, pages 995–1003, Berlin, Heidelberg, 2005. Springer Berlin Heidelberg.

[28] Indah Tiyas, Ali Barakbah, Tri Harsono, and Amang Sudarsono. Reinforced intrusion detection using pursuit reinforcement competitive learning. *EMITTER International Journal of Engineering Technology*, 2(1):39–49, 2014.

[29] James Cannady Georgia. Next generation intrusion detection: Autonomous reinforcement learning of network attacks. In *In Proceedings of the 23rd National Information Systems Secuity Conference*, pages 1–12, 2000.

[30] Arturo Servin. Towards traffic anomaly detection via reinforcement learning and data flow. pages 81–88.

[31] Fatma Belabed and Ridha Bouallegue. An optimized weight-based clustering algorithm in wireless sensor networks. *2016 International Wireless Communications and Mobile Computing Conference (IWCMC)*, 2016.

[32] Wei Zhang, Sajal Das, and Yonghe Liu. A trust based framework for secure data aggregation in wireless sensor networks. *IEEE Communications Society on Sensor and Ad Hoc Communications and Networks*, 2006.

[33] S. Seo, S. Park, and J. Kim. Improvement of network intrusion detection accuracy by using restricted boltzmann machine. In *8th International Conference on Computational Intelligence and Communication Networks (CICN)*, pages 413–417, Dec 2016.

[34] Daoying Ma and Aidong Zhang. An adaptive density-based clustering algorithm for spatial database with noise. *IEEE Intl Conf on Data Mining (ICDM'04)*.

[35] A. Ram, A. Sharma, A. S. Jalal, A. Agrawal, and R. Singh. An enhanced density based spatial clustering of applications with noise. In *IEEE Intl. Advance Computing Conference*, pages 1475–1478, March 2009.

[36] Random forests, leo breiman and adele cutler.

[37] Jiong Zhang, M. Zulkernine, and A. Haque. Random-forests-based network intrusion detection systems. *IEEE Trans. on Systems, Man, and Cybernetics, Part C*, 38/5:649–659, 2008.

[38] M.f Jiang, S.s Tseng, and C.m Su. Two-phase clustering process for outliers detection. *Pattern Recognition Letters*, 22/6-7:691–700, 2001.

[39] Daoying Ma and Aidong Zhang. An adaptive density-based clustering algorithm for spatial database with noise. *IEEE Intl Conf on Data Mining (ICDM'04)*.

[40] S. Doltsinis, P. Ferreira, and N. Lohse. An mdp model-based reinforcement learning approach for production station ramp-up optimization: Q-learning analysis. *IEEE Transactions on Systems, Man, and Cybernetics: Systems*, 44(9):1125–1138, Sept 2014.

[41] Christopher J. C. H. Watkins and Peter Dayan. Q-learning. *Machine Learning*, 8(3):279–292, May 1992.

[42] Xin Du and Jinjian Zhai. Algorithm trading using q-learning and recurrent reinforcement learning.

[43] Chris Gaskett, David Wettergreen, and Alexander Zelinsky. Q-learning in continuous state and action spaces. In Norman Foo, editor, *Advanced Topics in Artificial Intelligence*, pages 417–428, Berlin, Heidelberg, 1999. Springer Berlin Heidelberg.

[44] D. Kumar, N. Logganathan, and V. P. Kafle. Double sarsa based machine learning to improve quality of video streaming over http through wireless networks. In *2018 ITU Kaleidoscope: Machine Learning for a 5G Future (ITU K)*, pages 1–8, Nov 2018.